\documentclass[smallextended,numbook]{svjour3}

\journalname{Journal of Statistical Physics}
\journalname{}

\usepackage{ulem}

\usepackage{graphicx,float,amsmath,amssymb,mathrsfs}

\usepackage[english,francais]{babel}

\usepackage{url,hyperref}  

\usepackage[usenames,dvipsnames]{color}

\def\XXint#1#2#3{{\setbox0=\hbox{$#1{#2#3}{\int}$}
\vcenter{\hbox{$#2#3$}}\kern-.5\wd0}}

\newcommand{\ket}[1]{|\kern.3ex#1\kern.3ex\rangle}
\newcommand{\bra}[1]{\langle\kern.3ex #1 \kern.3ex|}

\newcommand{\mean}[1]{\left\langle #1\right\rangle}
\newcommand{\smean}[1]{\langle #1\rangle}
\newcommand{\EXP}[1]{e^{#1}}         

\newcommand{\sign}{\mathop{\mathrm{sign}}\nolimits}  

\newcommand{\heaviside}{\mathop{\theta_\mathrm{H}}\nolimits}  

\def\I{{\rm i}}
\newcommand{\deriv}[2]{\frac{\mathrm{d}#1}{\mathrm{d}#2}}

\def\D{{\rm d}}                  

\def\mass{m}
\def\clp{\gamma}
\def\perpar{\alpha}
\def\cfmat{\Gamma}

\begin{document}
\renewcommand{\labelitemi}{$\bullet$}
\renewcommand{\labelitemii}{$\star$}

\selectlanguage{english}

\title{Fluctuations of random matrix products and 1D Dirac equation with random mass}

\author{Kabir Ramola \and Christophe Texier}

\institute{
K. Ramola
\at 
Univ. Paris Sud ; CNRS ; Laboratoire de Physique Th\'eorique et Mod\`eles Statistiques, UMR 8626 ; 91405 Orsay cedex, France\\
\email{kabir.ramola@u-psud.fr}
\and
C. Texier
\at
Univ. Paris Sud ; CNRS ; Laboratoire de Physique Th\'eorique et Mod\`eles Statistiques, UMR 8626 ; 91405 Orsay cedex, France\\
\email{christophe.texier@u-psud.fr}
}

\date{July 10, 2014}

\maketitle

\begin{abstract}
We study the fluctuations of certain random matrix products $\Pi_N=M_N\cdots M_2M_1$ of $\mathrm{SL}(2,\mathbb{R})$, describing localisation properties of the one-dimen\-sional Dirac equation with random mass.
In the continuum limit, i.e. when matrices $M_n$'s are close to the identity matrix,
we obtain convenient integral representations for the variance $\Gamma_2=\lim_{N\to\infty}\mathrm{Var}(\ln||\Pi_N||)/N$.
The case studied exhibits a saturation of the variance at low energy $\varepsilon$ along with a vanishing Lyapunov exponent $\Gamma_1=\lim_{N\to\infty}\ln||\Pi_N||/N$, leading to the behaviour 
$\Gamma_2/\Gamma_1\sim\ln(1/|\varepsilon|)\to\infty$ as $\varepsilon\to0$.
Our continuum description sheds new light on the Kappus-Wegner (band center) anomaly.
\end{abstract}

\vspace{0.25cm}




\section{Introduction}

\begin{figure}[!ht]
\hspace{2cm}
\includegraphics[width=0.65\textwidth]{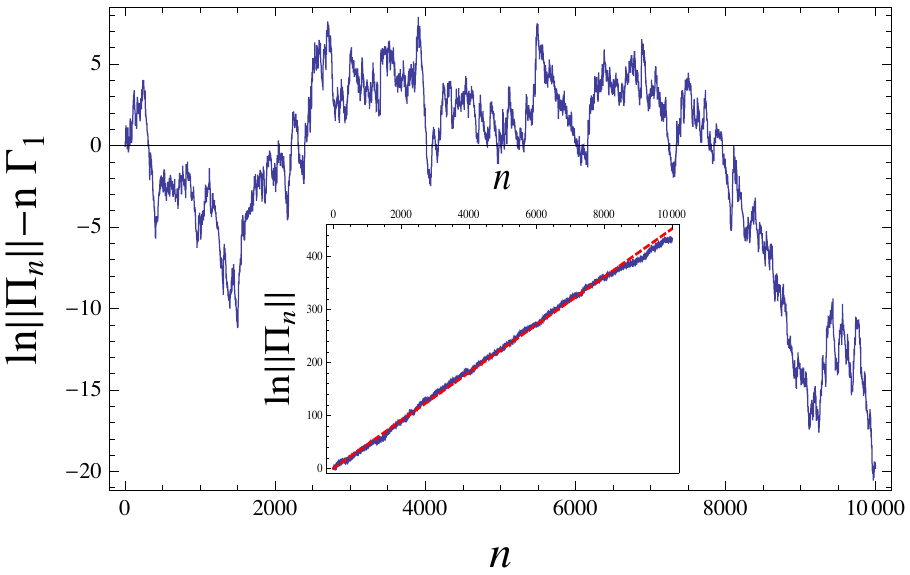}
\caption{{\it (color online) 
Fluctuations of a particular sequence of matrix products $\Pi_n=M_n\cdots M_1$, with matrices of the form \eqref{eq:MSusy1}.}}
\label{fig:FluctRMP}
\end{figure}

Transfer matrices lead to a convenient formulation of many statistical physics problems and have been extensively used 
since their introduction in the context of the Ising model \cite{Bax82}.
In the presence of randomness, most of the physics is captured by the Lyapunov exponent $\cfmat_1$ which quantifies the growth
rate of the matrix elements of a random matrix product (RMP) $\Pi_N=M_N\cdots M_2M_1$. 
Given the measure characterizing the independent and identically distributed random matrices $M_n$'s, 
the Furstenberg formula allows one to obtain, at least in principle, the Lyapunov exponent 
$\cfmat_1=\lim_{N\to\infty}\ln||\Pi_N||/N$, where $||\cdot||$ is a suitable norm, in terms of the solution of the Furstenberg's integral equation \cite{BouLac85}.
Besides the Lyapunov exponent, which describes the \textit{mean} free energy of the random 
Ising model 
\cite{Luc92,CriPalVul93,FigMosKno98}, the \textit{fluctuations} of RMP (Fig.~\ref{fig:FluctRMP}) 
also play an important role and are the main subject of this paper.
The study of fluctuations is related to the generalised Lyapunov exponent analysis 
\cite{Luc92,CriPalVul93}\,\footnote{Generalised central limit theorems for matrices was discussed in the mathematical literature (chapter V of the monograph~\cite{BouLac85}).
} 
and the multifractal formalism introduced by Paladin and Vulpiani \cite{PalVul87b,PalVul87}.
Fluctuations are of particular importance in the context of quantum localisation, where they dominate several physical quantities,
like the local density of states \cite{AltPri89} or the Wigner time delay \cite{TexCom99}.
Their precise characterisation is an important issue at the heart of the scaling approach used in the  justification of 
the single parameter scaling (SPS) hypothesis (cf. Refs.~\cite{AndThoAbrFis80,CohRotSha88} and references therein). 
In the last decade, this question has been re-examined more precisely for a lattice model \cite{DeyLisAlt00,SchTit03,TitSch05}. 
Recently Lyapunov exponents have been analytically obtained for general RMPs of $\mathrm{SL}(2,\mathbb{R})$ in the \textit{continuum limit} \cite{ComLucTexTou13}. This has significantly improved our understanding of RMPs and of one-dimensional (1D) quantum localisation models due to their close connection \cite{ComTexTou10}.
The present work is a first step towards generalising this approach for the fluctuations.
We will consider matrices belonging to  two particular subgroups of 
$\mathrm{SL}(2,\mathbb{R})$~:
\begin{equation}
  \label{eq:MSusy1}
  M_n = 
  \left(
  \begin{array}{cc}
    \cos \theta_n & -\sin \theta_n \\
    \sin \theta_n & \phantom{-}\cos \theta_n 
  \end{array}    
  \right)
  \left(
  \begin{array}{cc}
    \EXP{\eta_n} & 0 \\
    0 & \EXP{-\eta_n}
  \end{array}    
  \right)
\end{equation}
or
\begin{equation}
  \label{eq:MSusy2}
  M_n = 
  \left(
  \begin{array}{cc}
    \cosh \tilde\theta_n & \sinh \tilde\theta_n \\
    \sinh \tilde\theta_n & \cosh \tilde\theta_n 
  \end{array}    
  \right)
  \left(
  \begin{array}{cc}
    \EXP{\eta_n} & 0 \\
    0 & \EXP{-\eta_n}
  \end{array}    
  \right)
  \:.
\end{equation}
We will show in section~\ref{sec:mapping} that products of matrices of the type \eqref{eq:MSusy1} are transfer matrices for the bi-spinor $\Psi=(\psi,\chi)$ solution of the 1D Dirac equation
\begin{equation}
  \label{eq:Dirac}
  \left[ \sigma_2\,\I\partial_x + \sigma_1\, \mass(x) \right]\Psi(x) = \varepsilon\,\Psi(x)
\end{equation}
for a mass of the form $\mass(x)=\sum_n\eta_n\,\delta(x-x_n)$ and with $\theta_n=\varepsilon\,(x_{n+1}-x_n)\in\mathbb{R}$, where $\sigma_i$'s are Pauli matrices. 
Matrices of the type \eqref{eq:MSusy2} with $\tilde\theta_n=-\I\varepsilon\,(x_{n+1}-x_n)$ correspond to the case $\varepsilon\in\I\mathbb{R}$. 
The Dirac equation \eqref{eq:Dirac} with random mass is a relevant model in several contexts of condensed matter, e.g.
random spin chains or organic conductors (see references in Refs.~\cite{LeDMonFis99,TexHag10}). It can also be exactly mapped onto supersymmetric quantum mechanics \cite{ComTex98} and the 
Sinai problem of 1D classical diffusion in a random force field \cite{BouComGeoLeD90,LeDMonFis99}.
Many properties of the model can be obtained exactly when the mass is chosen to be a Gaussian white noise~: 
\begin{equation}
  \mean{\mass(x)}=\mu\,g 
  \mbox{ and }
  \mean{\mass(x)\mass(x')}_c=g\,\delta(x-x')
  \:,
\end{equation}
where $\mean{XY}_c=\mean{XY}-\mean{X}\mean{Y}$. 
For example the Lyapunov exponent, defined in the localisation problem as $\clp_1 = \lim_{x\to\infty}{\ln|\psi(x)|}/{x}$, is known~\cite{BouComGeoLeD90}
\begin{equation}
\label{eq:BouchaudComtetGeorgeLedoussal}
  \clp_1 = -\mu g
  + \varepsilon\,\frac{H^{(1)}_{\mu+1}(\varepsilon/g)}{H^{(1)}_{\mu}(\varepsilon/g)}
  \:,
\end{equation}
where $H^{(1)}_{\mu}(z)$ is the Hankel function. 
The case $\mean{\mass(x)}=0$ is of particular interest since the Lyapunov exponent vanishes as $\varepsilon\to0$, indicating a \textit{delocalisation} point in the spectrum.
In this unusual case, the characterisation of 
$
  \clp_2  = \lim_{x\to\infty}\mathrm{Var}(\ln|\psi(x)|)/x
$
(i.e. $\cfmat_2=\lim_{N\to\infty}\mathrm{Var}(\ln||\Pi_N||)/N$) is thus crucial. We will show that the 
fluctuations saturate as~$\varepsilon\to0$, and thus dominate localisation properties.

\section{Mapping}
\label{sec:mapping}

The mapping between RMP and 1D localisation models like the random Kronig-Penney model \cite{Luc92}, 
was recently extended to general RMPs of $\mathrm{SL}(2,\mathbb{R})$ \cite{ComTexTou10}.
For the case of interest here, the mapping works as follows~: consider a random mass given as a superposition of delta-functions $\mass(x)=\sum_n\eta_n\,\delta(x-x_n)$, 
where coordinates are ordered $x_1<x_2<\cdots$.
Matching conditions across each impurity read 
$\psi(x_n^+)=\psi(x_n^-)\EXP{\eta_n}$ and $\chi(x_n^+)=\chi(x_n^-)\EXP{-\eta_n}$, hence the diagonal matrix in \eqref{eq:MSusy1}, while the rotation of angle $\theta_n=\varepsilon\,(x_{n+1}-x_{n})$ stands for the free evolution between two impurities.
If we consider the Dirac equation with a purely imaginary energy $\varepsilon\in\I\mathbb{R}$, the matrix \eqref{eq:MSusy2} with $\tilde\theta_n=-\I\theta_n\in\mathbb{R}$, relates 
$(\psi,\tilde\chi)=(\psi,-\I\chi)$ at $x_n^-$ and $x_{n+1}^-$.~\footnote{Note that matrices \eqref{eq:MSusy2} with $\tanh\tilde{\theta}_n=\EXP{-2\beta J_n}$ and $\eta_n=\beta h_n$ are tranfer matrices for the random Ising chain with couplings $J_n$ and magnetic fields $h_n$ \cite{Luc92}~;
a continuum approximation of the model was considered in Ref.~\cite{FigMosKno98} allowing these authors to recover the Lyapunov exponent \eqref{eq:BouchaudComtetGeorgeLedoussal} obtained first in~Ref.~\cite{BouComGeoLeD90}.
}
The product $\Pi_N=M_N\cdots M_2M_1$ thus controls the value of the spinor and the study of the growth of the RMP characterizes the localisation properties of the wave function. 
It is convenient to introduce the Riccati variable $z(x)=-\varepsilon\,\chi(x)/\psi(x)$~; from Eq.~\eqref{eq:Dirac}, we find
\begin{equation}
  \label{eq:SDEsusy}
  \deriv{}{x}z(x)=-\varepsilon^2-z(x)^2-2z(x)\,\mass(x)
  \:.
\end{equation} 
If the lengths $\ell_n=x_{n+1}-x_n>0$ are either equal (lattice) or distributed with an exponential law $P(\ell)=\rho\,\EXP{-\rho\ell}$, the stochastic differential equation (SDE) defines a Markov process.
Hence, $\Psi(x_{N+1}^-)=\Pi_N\Psi(x_{1}^-)$ shows that $\ln||\Pi_N||$ and $\ln|\psi(x)|$ are asymptotically equivalent, thus their cumulants are related by $\clp_n=\rho\,\cfmat_n$. 
In the following we will consider the continuum limit of the RMP problem when the random parameters 
are small $\theta_n=\varepsilon\ell_n\to0$ and $\eta_n\to0$, i.e. the matrices $M_n$ are close to the identity matrix, 
in such a way that $\mean{\eta_n}=0$ and $g=\mean{\eta_n^2}/\mean{\ell_n}$ is fixed~; 
this limit corresponds to the case where $\mass(x)$ is a Gaussian white noise with zero mean \cite{ComLucTexTou13,ComTexTou13}. 
The SDE \eqref{eq:SDEsusy} must be interpreted in the Stratonovich convention as is usual in physical problems~\cite{Gar89}. 
The study of the fluctuations of the RMP can be performed by introducing the generalised Lyapunov exponent~\cite{PalVul87,CriPalVul93}
\begin{equation}
  \label{eq:DefGeneralisedLyapunov}
  \Lambda(q) = \lim_{x\to\infty}\frac{\ln\mean{|\psi(x)|^q}}{x}
  =\sum_{n=1}^\infty\frac{q^n}{n!}\clp_n
  \:,
\end{equation}
which is the generating function for the cumulants of $\ln|\psi(x)|$.
In the following discussion we focus on $\clp_2$. 
From the definition of the Riccati variable, we may write $\ln|\psi(x)|=\int_0^x\D t\,\big[z(t)+\mass(t)\big]$, hence
\begin{equation}
  \label{eq:ConvenientKappa2}
  \clp_2 = g
  + 2\lim_{x\to\infty}\int_0^x\D t\,
  \mean{z(x)\left[z(t)+\mass(t)\right]}_c
  \:.
\end{equation}
It is convenient to use the relation
\begin{equation}
  \label{eq:UsefulRelation}
  2\int_{x_0}^x\D t\,\left[z(t)+\mass(t)\right]
  = - \ln\left|\frac{z(x)}{z(x_0)}\right|
  +   \int_{x_0}^x\D t\, \left( z(t) - \frac{\varepsilon^2}{z(t)} \right)
  \:,
\end{equation}
obtained by integration of \eqref{eq:SDEsusy}.
Finally we get
\begin{equation}
  \label{eq:TheResultSusy}
      \clp_2 = g
  - \mean{z \, \ln|z/\varepsilon|}
   +\int\D z\D z'\,
  z\, G(z|z') \, \left(z' - \frac{\varepsilon^2}{z'} \right)\,f(z')
  \:.
\end{equation}
The propagator is defined as 
\begin{equation}
  G(z|z')=\int_0^\infty\D x\left[P_x(z|z')-f(z)\right]
\end{equation}
where $P_x(z|z')$ is the conditional probability, solution of the Fokker-Planck equation $\partial_xP_x(z|z')=\mathscr{G}^\dagger P_x(z|z')$, and $f(z)=\lim_{x\to\infty}P_x(z|z')$ is the stationary distribution of the Riccati variable, with 
\begin{equation}
\mathscr{G}^\dagger=2g\,\partial_zz\partial_zz+\partial_z(z^2+\varepsilon^2)
\end{equation}
being the forward generator of the diffusion (adjoint of the generator).
Eq.~\eqref{eq:TheResultSusy} is one of our main results~: $f$ can be explicitly obtained as the normalisable solution of $\mathscr{G}^\dagger f=0$ and $G$ solves 
\begin{equation}
\mathscr{G}^\dagger G(z|z')=f(z)-\delta(z-z')
\:.
\end{equation}
Note that, in the derivation of the second term of \eqref{eq:TheResultSusy}, we have used the underlying supersymmetry of the Dirac equation \cite{ComTexTou10,ComTexTou13} $f(z)=f(-\varepsilon^2/z)\,|\varepsilon|^2/z^2$, leading to $\mean{\ln|z|}=\ln|\varepsilon|$.
Solving the equation for $G$, we can obtain an explicit representation for $\clp_2$ in terms of multiple integrals, like it is done for another subgroup of $\mathrm{SL}(2,\mathbb{R})$ in Appendix~\ref{appendix:Schrod}. 
We prefer to proceed in a different manner in order to derive limiting values for~$\clp_2$.

\section{Universal regime (large real energy $\varepsilon\gg g$)}

The large energy limit is the \textit{universal} regime where SPS holds \cite{CohRotSha88}~: a unique scale controls the average and the fluctuations $\clp_2\simeq\clp_1$.
The variance was explicitly calculated for this model in Ref.~\cite{Tex99} and coincides with the known value for the Lyapunov exponent \cite{BouComGeoLeD90}, 
Eq.~\eqref{eq:BouchaudComtetGeorgeLedoussal}, 
that saturates at high energy~: $\clp_2\simeq g/2$ (Fig.~\ref{fig:gamma2}).

\section{Small real energy $\varepsilon\ll g$}

The process $z(x)$ flows through the full interval $\mathbb{R}$ and it is convenient to consider the variable 
\begin{equation}
\zeta=\mp\ln(\pm z/|\varepsilon|)/2
  \hspace{0.25cm}\mbox{ for }
z\in\mathbb{R}_\pm
\end{equation}
When $z(x)$ goes from $+\infty$ to $0$ the process $\zeta(x)$ crosses $\mathbb{R}$ once, and a second time when $z(x)$ goes from $0$ to $-\infty$.
The new process obeys the SDE
\begin{equation}
  \label{eq:LangevinZeta}
  \deriv{}{x}\zeta(x) = -\mathcal{U}'(\zeta(x)) + \mass(x)
\end{equation}
for the unbounded potential 
\begin{equation}
\mathcal{U}(\zeta) = - \frac{|\varepsilon|}{2}\,\sinh2\zeta
\:.
\end{equation}
Rewriting \eqref{eq:TheResultSusy} in terms of the new variable, we get 
\begin{equation}
  \label{eq:Gamma2Susy}
  \clp_2 = g - 2\mean{\zeta\,\mathcal{U}'(\zeta)}
  + 8
  \int\D\zeta\D\zeta'\,
     \mathcal{U}(\zeta)\,  \mathcal{G}(\zeta|\zeta')\, \mathcal{U}(\zeta') \,\mathcal{P}(\zeta')
\end{equation}
where $\mathscr{G}^\dagger\mathcal{P}=0$ and  
\begin{equation}
\label{eq:EqForGsusy}
\mathscr{G}^\dagger \mathcal{G}(\zeta|\zeta') = \mathcal{P}(\zeta)-\delta(\zeta-\zeta')
\end{equation}
for the forward generator 
\begin{equation}
\mathscr{G}^\dagger = \frac{g}{2}\,\partial_\zeta^2 + \partial_\zeta\mathcal{U}'(\zeta)
\:.
\end{equation}
The details of the derivation of Eq.~\eqref{eq:Gamma2Susy} are given in Appendix~\ref{appendix:DetailsOnCentralEq}.
The variable $\zeta$ is appropriate for the low energy analysis~: the exponential dependence of the potential clearly illustrates 
the decoupling between the ``deterministic force'' $\mathcal{U}'(\zeta)$ and Langevin ``force'' $\mass(x)$. 
We can map the problem onto an effective free diffusion problem in the interval 
$[\zeta_-,\zeta_+]$, where $\zeta_\pm=\pm\ln(2g/|\varepsilon|)/2$.
The form of $\mathcal{U}(\zeta)$ at infinity leads to the boundary conditions~:
absorption at one boundary, $\mathcal{P}(\zeta_+)=0$, 
with reinjection of the current at the other boundary, $\mathcal{P}'(\zeta_-)=\mathcal{P}'(\zeta_+)$.
The stationary distribution takes the approximate form 
\begin{equation}
  \mathcal{P}(\zeta) \simeq
  \frac{2(\zeta_+-\zeta)}{(\zeta_+-\zeta_-)^2}
  \hspace{0.5cm}\mbox{for } \zeta\in[\zeta_-,\zeta_+]
\end{equation}
and the solution of Eq.~\eqref{eq:EqForGsusy} is given by
\begin{align}
  \mathcal{G}(\zeta|\zeta')
  \simeq 
  \frac{2}{g}
  \bigg\{
  &- \frac{1}{6}(\zeta_+-\zeta)
  + \frac{1}{3(\zeta_+-\zeta_-)^2}\Big[(\zeta_+-\zeta_>)^3
  \nonumber\\
  \label{eq:GForRealEnergy}
  &+3(\zeta_<-\zeta_-)^2(\zeta_+-\zeta_>)+\heaviside(\zeta'-\zeta)\,(\zeta'-\zeta)^3
  \Big]
  \bigg\}
  \:,
\end{align}
where $\heaviside(\zeta)$ is the Heaviside function, $\zeta_>=\mathrm{max}(\zeta,\zeta')$ and $\zeta_<=\mathrm{min}(\zeta,\zeta')$. 
As a check, we recover 
that the Lyapunov exponent 
\begin{align}
  \label{eq:EpressionLyapunov}
  \clp_1=2\mean{\mathcal{U}(\zeta)}
\end{align}
\textit{vanishes} as 
\begin{align}
  \label{eq:LimitLyapunov}
  \clp_1\underset{|\varepsilon|\to0}{\simeq} \frac{g}{\ln(2g/|\varepsilon|)}
  \:,
\end{align}
a behaviour which coincides
with the asymptotic of the exact result \eqref{eq:BouchaudComtetGeorgeLedoussal}. 
We easily compute \eqref{eq:Gamma2Susy}, leading to
\begin{equation}
  \label{eq:LimitSmallEnergy1}
  \clp_2 \underset{\varepsilon\to0}{\simeq}
  g\,
  \left[
    \frac{1}{3}
    +\frac{1}{2\ln(2g/|\varepsilon|)}
  \right]
  \:,
\end{equation}
which shows the \textit{saturation} of the fluctuations as $\varepsilon\to0$.

\section{Small complex energy $-\I\varepsilon\ll g$}

\begin{figure}[!ht]
\hspace{1.5cm}
\includegraphics[width=0.75\textwidth]{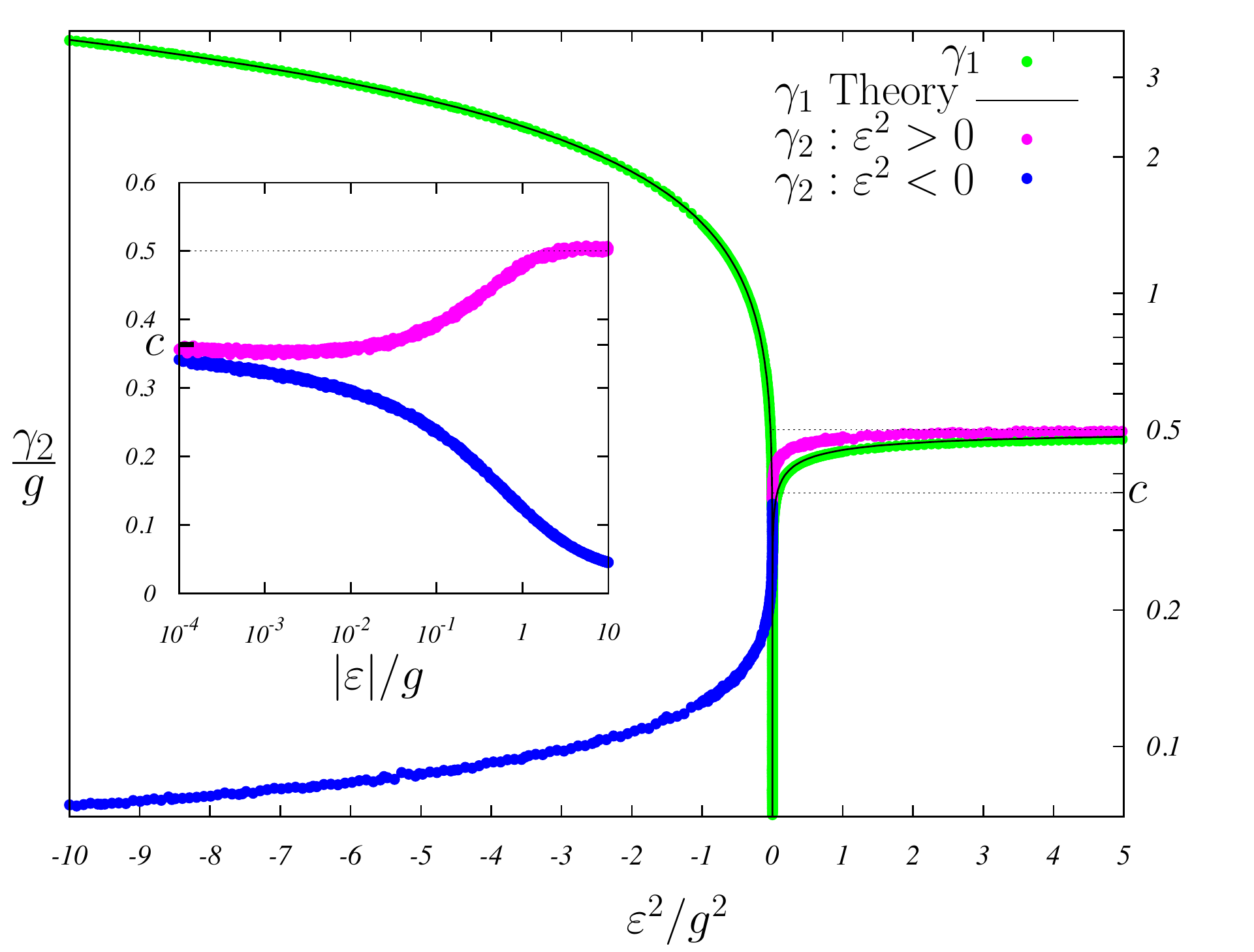}
\caption{\it (color online) Plot of the Lyapunov exponent ($\clp_1$) and the variance ($\clp_2$). 
The solid black line corresponds to the exact result \eqref{eq:BouchaudComtetGeorgeLedoussal}.
For $|\varepsilon|\to0$, $\clp_1$ vanishes while the variance saturates to $c=1-{2}/{\pi}$.
Inset~: Plot in log-linear scale showing the low energy saturation of $\clp_2$.}
\label{fig:gamma2}
\end{figure}

For complex energy $\varepsilon\in\I\mathbb{R}$, the process $z(x)$ is trapped on $\mathbb{R}_+$. 
The SDE \eqref{eq:LangevinZeta} still holds for the bounded potential 
\begin{equation}
\mathcal{U}(\zeta) = \frac{|\varepsilon|}{2}\,\cosh2\zeta
\:.
\end{equation}
Making use of the fact that $\mathcal{U}(\zeta)$ is symmetric, we can show that the representations \eqref{eq:Gamma2Susy} and \eqref{eq:EpressionLyapunov} are still valid 
(see Appendix~\ref{appendix:DetailsOnCentralEq}),
the stationary distribution being now an equilibrium distribution $\mathcal{P}(\zeta)\propto\exp\big[-2\,\mathcal{U}(\zeta)/g\big]$.
In the low energy limit, we again use the decoupling between the deterministic force and the Langevin force~: the effect of the confining potential is now replaced by reflecting boundary conditions $\mathcal{P}'(\zeta_-)=\mathcal{P}'(\zeta_+)=0$.
The stationary distribution is 
\begin{equation}
  \mathcal{P}(\zeta)
  \simeq
  \frac{1}{\zeta_+-\zeta_-}
  \hspace{0.5cm}\mbox{for } \zeta\in[\zeta_-,\zeta_+]
  \:,
\end{equation}
thus \eqref{eq:EpressionLyapunov} again leads to \eqref{eq:LimitLyapunov}.
The propagator takes the form
\begin{align}
  \label{eq:GForComplexEnergy}
   \hspace{-0.35cm}
  \mathcal{G}(\zeta|\zeta') \simeq 
  \frac2g\,
  \left\{
    -\frac{|\zeta-\zeta'|}{2} + \frac{\zeta^2+\zeta'^2}{2(\zeta_+-\zeta_-)} 
    + \frac{\zeta_+-\zeta_-}{12}
  \right\}
\end{align}
and Eq.~\eqref{eq:Gamma2Susy} gives
\begin{equation}
  \label{eq:LimitSmallEnergy2}
  \clp_2 \underset{\varepsilon\to\I 0}{\simeq}
  g\,
  \left[
    \frac{1}{3}
    -\frac{1}{2\ln(2g/|\varepsilon|)}
  \right]
  \:.
\end{equation}
This shows that, as a function of $\varepsilon^2$, the variance $\clp_2$ is continuous around $0$.
Setting $\varepsilon=0$, a more direct analysis can be performed (Appendix~\ref{appendix:FlucZero}) showing that  
\begin{equation}
  \clp_2 = g\left(1-\frac{2}{\pi}\right)\simeq0.363\,g
  \hspace{0.25cm}\mbox{ for } 
  \varepsilon=0
  \:.
\end{equation}
The small discrepancy ($\sim8\%$) is explained by the fact that the constant term in (\ref{eq:LimitSmallEnergy1},\ref{eq:LimitSmallEnergy2}) is sensitive to the precise position of the cutoffs $\zeta_\pm$ of the free diffusion approximation.
Numerical calculations confirm the saturation and suggest a logarithmic behaviour, although it is difficult to precisely fit this logarithmic correction (Fig.~\ref{fig:gamma2}).

\section{Large complex energy $-\I\varepsilon\gg g$~: A perturbative treatment of the stochastic differential equation}

In this case, it is convenient to develop a perturbative approach based on the SDE~\eqref{eq:SDEsusy}. 
We perform the rescaling 
\begin{equation}
\label{eq:rescaled_process}
  z(x)=|\varepsilon|+\sqrt{g|\varepsilon|}\,y(u)
  \hspace{0.5cm}\mbox{with}\hspace{0.5cm}
  x=u/|\varepsilon|
  \:,
\end{equation}
leading to 
\begin{equation}
    \deriv{y(u)}{u} = -2\, y(u) - 2\,\eta(u)
  -\perpar\, \left[ y(u)^2 + 2\,y(u)\,\eta(u) \right]
  \:,
\end{equation}
where $\eta(u)$ is a normalised Gaussian white noise, $\mean{\eta(u)\eta(u')}=\delta(u-u')$. 
The perturbative parameter is $\perpar = \sqrt{{g}/{|\varepsilon|}}$.
Expansion of the process in powers of $\perpar$, as $y=y_0+y_1+y_2+\cdots$, leads to the explicit integral representations
\begin{align}
  y_0(u) &=  - 2 \int_0^u\D t\,\EXP{-2(u-t)}\, \eta(t)
  \:,\\
  y_1(u) &= -\perpar
   \int_0^u\D t\,\EXP{-2(u-t)}\, \left[y_0(t)^2+2y_0(t)\eta(t)\right]
   \:,\\
  y_2(u) &= -2\perpar
   \int_0^u\D t\,\EXP{-2(u-t)}\, \left[y_0(t)+\eta(t)\right]y_1(t)
   \:,
\end{align}
where transient terms have been neglected. Order zero $y_0(u)$ is the Ornstein-Uhlenbeck process.
These expressions are suitable for computing $\clp_1=\mean{z}\simeq|\varepsilon|+\sqrt{g|\varepsilon|}\,\mean{y_1(u)}=|\varepsilon|+g/2$ and the correlator \eqref{eq:ConvenientKappa2}.
The latter can be rewritten in terms of the rescaled process as
\begin{equation}
\clp_2 = g \left \{1 + 2 \int_{0}^{\infty} \D u\, \langle y(u_0) y(u + u_0) \rangle_c + 2 \langle y(u_0) \int_{0}^{u_0} \D u\, \eta(u) \rangle_c \right \},
\end{equation}
where $u_0 \to \infty$.
It is easy to see that the first non-zero contribution to this expression comes at order $\perpar^2$.
We then have the following expression for $\clp_2$ to lowest order in $\perpar$
\begin{align}
\label{eq:gamma_2_secondorder}
\nonumber
\clp_2 = 2 g &\left\{ 
       \int_{0}^{\infty} \D u\, 
         \Big[ 
            \mean{ y_0(u_0) y_2(u + u_0) }_c 
            +\mean{ y_2(u_0) y_0(u + u_0) }_c 
       \right.\\
           & + \mean{ y_1(u_0) y_1(u + u_0) }_c
         \Big] 
\left. + \smean{ y_2(u_0) \int_{0}^{u_0} \D u\, \eta(u) }_c + \mathcal{O}(\perpar^4)\right\}
\:.
\end{align}
It is possible to compute this expression exactly. Since the process is derived from a Gaussian white noise source,
we can use Wick's theorem to reduce all the correlation functions to products over two-point correlation functions. However,
the full calculation is rather cumbersome. Instead, we use the stochastic calculus functionalities of {\tt Mathematica 9.0}
to derive the values of the correlators. We have
\begin{align}
\lim_{u_0\to\infty}
\int_{0}^{\infty} \D u \,
\big[\mean{ y_0(u_0) y_2(u + u_0)}_c + \mean{y_2(u_0) y_0(u + u_0) }_c \big] &= \frac{3}{4} \perpar^2 
\:,
\\
\lim_{u_0\to\infty}
\int_{0}^{\infty} \D u \,\langle y_1(u_0) y_1(u + u_0) \rangle_c &= \frac{1}{8} \perpar^2
\:,
\\
\lim_{u_0\to\infty}
\langle y_2(u_0) \int_{0}^{u_0} \D u\, \eta(u) \rangle_c &= -\frac{3}{4} \perpar^2.
\end{align}
Summing these three contributions,
we arrive at
\begin{equation}
\label{eq:PowerLawDecay}
\clp_2 \underset{\varepsilon\to\I\infty}{\simeq} \frac{g^2}{4|\varepsilon|}
\:.
\end{equation}

\section{Numerical calculations}
\label{sec:numerics}

\subsection{Method}

We have performed a Monte Carlo simulation of the matrix problem, i.e. of the Dirac equation for the random mass
\begin{equation}
  \label{eq:RandomMass}
  \mass(x) = \sum_n\eta_n\,\delta(x-x_n)
  \:,
\end{equation}
where the impurities are independently and uniformly dropped on the line with a mean density $\rho$. This corresponds to an exponential distribution $P(\ell)=\rho\,\EXP{-\rho\ell}$ for the distance $\ell_n=x_{n+1}-x_n>0$ between consecutive impurities.
The mass is uncorrelated in space, i.e. is a \textit{non-Gaussian} white noise.
The limit of the Gaussian white noise considered in the previous sections corresponds to $\rho\to\infty$ and $\eta_n\to0$ with $\mean{\eta_n}=0$ and $g=\rho\mean{\eta_n^2}$ fixed.
This is a continuum model that is easy to implement numerically.

\paragraph{Real energy.--}

We parametrize the spinor as $\Psi=\EXP{\xi}(\sin\Theta,-\cos\Theta)$ and study the evolution of the two variables.
Between impurities $n$ and $n+1$ we have obviously
$\Theta_{n+1}^--\Theta_{n}^+=\varepsilon\ell_n$ and $\xi_{n+1}^--\xi_{n}^+=0$,
where we have introduced the notation 
$\Theta_n^\pm=\Theta(x_n^\pm)$ and $\xi_n^\pm=\xi(x_n^\pm)$.
Across the impurity \cite{BieTex08}
\begin{align}
 \tan\big(\Theta_{n}^+\big) &= \tan\big(\Theta_{n}^-\big)\, \EXP{2\eta_n}
 \\ 
 \xi_{n}^+ - \xi_n^-   
   &= \frac{1}{2}\, \ln\big[ \EXP{2\eta_n} \sin^2\Theta_n^- + \EXP{-2\eta_n} \cos^2\Theta_n^+  \big]
\:.
\end{align}
The norm of the RMP is identified with the norm of the spinor
\begin{equation}
  \ln||\Pi_N||_{\Psi_0} = \xi(x_{N+1}^-) = \frac{1}{2}\ln\left[\Psi(x_{N+1}^-)^\dagger\Psi(x_{N+1}^-)\right]
  \:.
\end{equation}

\paragraph{Complex energy.--}

For $\varepsilon\in\I\mathbb{R}$, we write the spinor as 
$\Psi=\EXP{\xi}(\sin\Theta,-\I\cos\Theta)$. 
Evolution of the two variables due to the rotation of complex angle is \cite{Tex99}
\begin{align}
 \tan\big(\Theta_{n+1}^-+{\pi}/{4}\big)
&= \tan\big(\Theta_{n}^++{\pi}/{4}\big)\, \EXP{2|\varepsilon|\ell_n}
\\
\xi_{n+1}^- - \xi_n^+ &= \frac{1}{2}\, \ln\big[{\cos2\Theta_{n}^+}/{\cos2\Theta_{n+1}^-} \big]
\:.
\end{align}
Evolution across an impurity are similar to the case of real energy.

\subsection{The saturation of $\clp_2$ for $\varepsilon\to\I\infty$ when $\mass(x)$ is a non-Gaussian white noise}
\label{subsec:saturation}

For large but finite density $\rho$ we show that the fluctuations saturate for large complex energy $|\varepsilon|\gg\rho$.
Expansion of the previous equations in the limit $|\varepsilon|\ell_n\to\infty$ gives
$
  \xi_{n+1}^- - \xi_n^-
  = |\varepsilon|\,\ell_n + \ln\cosh\eta_n + \mathcal{O}(\EXP{-2|\varepsilon|\ell_{n-1}}) + \mathcal{O}(\EXP{-2|\varepsilon|\ell_n})
$ .
We deduce the following representation for the process $\xi(x)$
\begin{equation}
    \xi(x) = |\varepsilon|x
  +\sum_{n=1}^{\mathscr{N}(x)} \ln\cosh\eta_n 
  +  \mathcal{O}\left(\frac{\rho}{|\varepsilon|}\rho x\right)
\end{equation}
where $\mathscr{N}(x)$ is the number of impurities on the interval $[0,x]$.
$\mathscr{N}(x)$ is a Poisson process and $\xi(x)$ a compound Poisson process (see for instance the introduction of Ref.~\cite{GraTexTou14} and references therein). 
Using standard properties of compound Poisson processes, we obtain
\begin{align}
  \clp_1 &= \lim_{x\to\infty}\frac{\xi(x)}{x}  \simeq |\varepsilon| + \rho \mean{\ln\cosh\eta_n }
  \\
  \clp_2 &= \lim_{x\to\infty}\frac{\mathrm{Var}(\xi(x))}{x} \simeq \rho \mean{\ln^2\cosh\eta_n }
\end{align}
(note that the \textit{cumulants} of $\xi(x)$ involves the \textit{moments} of the jump amplitudes).
Hence for $|\varepsilon|\gg\rho$ with $\eta_n\ll1$ we obtain 
$\clp_1\simeq |\varepsilon|+\rho\mean{\eta_n^2}/2=|\varepsilon|+g/2$ 
and $\clp_2 \simeq \rho\mean{\eta_n^4}/4$.

\begin{figure}[!ht]
\hspace{1.5cm}
\includegraphics[width=0.75\textwidth]{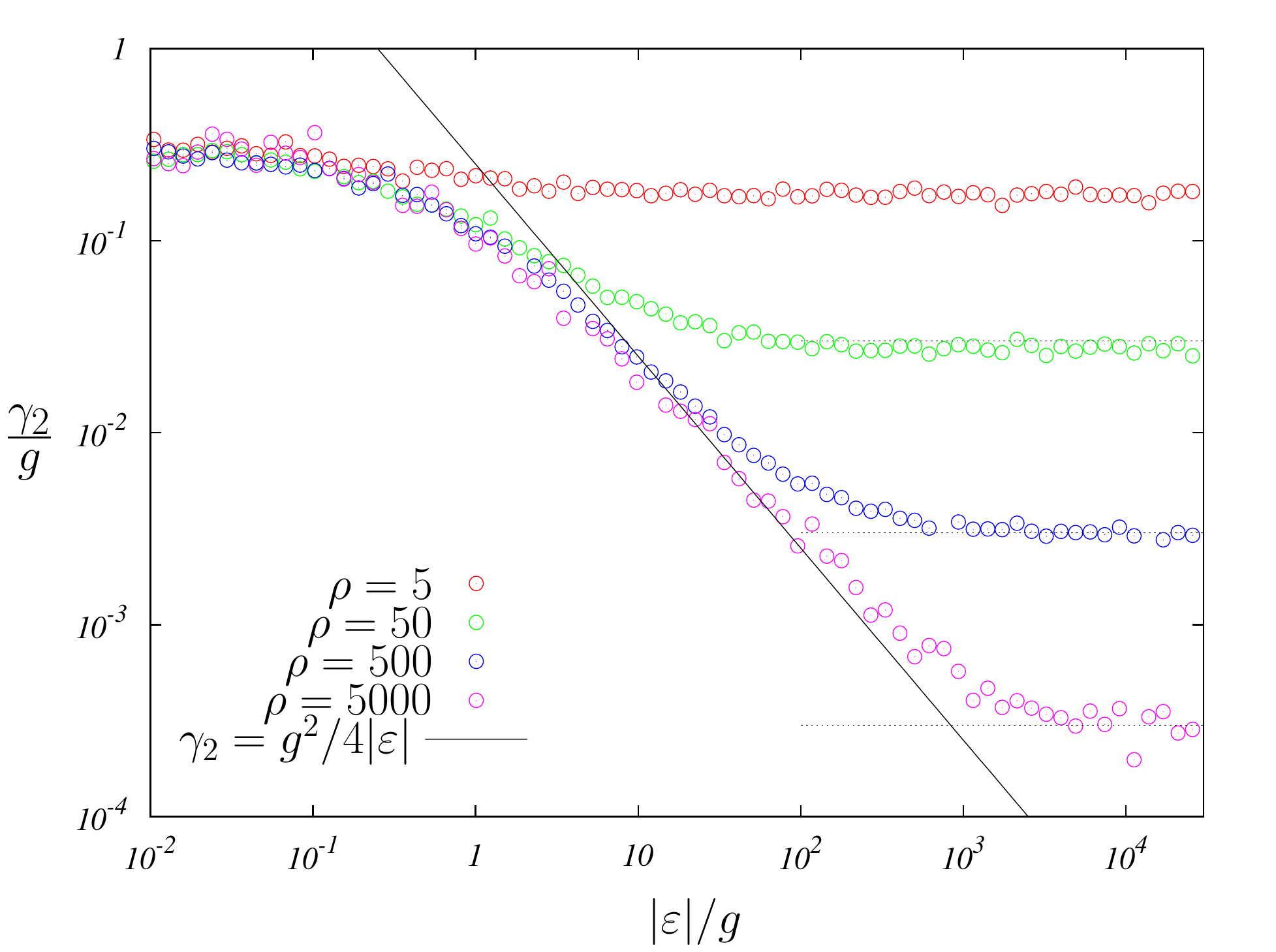}
\caption{\it (color online) Variance for matrices of type \eqref{eq:MSusy2} with exponentially distributed angles $\smean{\tilde\theta_n}=|\varepsilon|/\rho$ and weights distributed according to a symmetric exponential law with $g=\rho\mean{\eta_n^2}$. The saturation values are $\clp_2\simeq3g^2/2\rho$ (dotted lines).
The crossover between the two behaviours takes place at a scale $|\varepsilon|\sim\rho$.}
\label{fig:Gamma2LargeNeg}
\end{figure}

\subsection{Results}

As a first check, we compare the Lyapunov exponent $\clp_1$ obtained from the procedure explained above with the analytical expression \eqref{eq:BouchaudComtetGeorgeLedoussal}~: 
green dots and black continuous line on Fig.~\ref{fig:gamma2}, respectively. The agreement is excellent.

For large real energy we see on the figure that $\clp_2$ saturates at the same value as $\clp_1$ (SPS).
For small energy, $|\varepsilon|\to0$, the inset of Fig.~\ref{fig:gamma2} shows the logarithmic behaviours.

The behaviour \eqref{eq:PowerLawDecay} is more difficult to observe as it is a property of the Dirac equation when the mass $\mass(x)$ is a \textit{Gaussian} white noise.
The numerical simulation is performed for a \textit{non-Gaussian} white noise, Eq.~\eqref{eq:RandomMass}, which leads to a saturation of the fluctuations, as explained in paragraph~\ref{subsec:saturation}.
For this reason, the power law decay \eqref {eq:PowerLawDecay} is only obtained in an intermediate range of energy $g\ll|\varepsilon|\ll\rho$ (Fig.~\ref{fig:Gamma2LargeNeg}), and we observe the saturation for $|\varepsilon|\gg\rho$.
Choosing weights distributed according to a symmetric exponential law like in Ref.~\cite{ComTexTou13}, 
the saturation value is $\clp_2\simeq\rho\mean{\eta_n^4}/4=3g^2/(2\rho)$, in agreement with the numerics (black dotted lines in Fig.~\ref{fig:Gamma2LargeNeg}).

\section{Localisation}

\subsection{Low energy localisation}

The saturation of the fluctuations concomitant with the vanishing of the Lyapunov exponent has important consequences for 
the localisation. While the Lyapunov exponent is usually introduced as a measure of the localisation 
(see the monographs \cite{LifGrePas88,Luc92} or the review \cite{ComTexTou13}),
for a given small energy $|\varepsilon|\ll g$, the fluctuations dominate
$$
\sqrt{\clp_2x}\gtrsim\clp_1x
\hspace{0.5cm}\mbox{for}\hspace{0.5cm}
x\lesssim\xi_\varepsilon= (1/g)\ln^2(g/|\varepsilon|)
\:,
$$
i.e. on a scale $\xi_\varepsilon$ much larger than the inverse Lyapunov exponent $1/\clp_1\sim(1/g)\ln(g/|\varepsilon|)$.
The scale $\xi_\varepsilon$ has appeared in other studies~: 
in the average Green's function \cite{BouComGeoLeD90} (see discussion and references in Ref.~\cite{SteCheFabGog99}), 
in the distribution of the distances between consecutive nodes of the wave function \cite{TexHag10},
or in the boundary sensitive average local \cite{SteFabGog98} and global density of states \cite{TexHag10} (Thouless criterion).
This is a new indication that \textit{the Lyapunov exponent cannot be interpreted as the inverse localisation length in this case} \cite{TexHag10,ComTexTou13}.

\subsection{Band center anomaly}

\begin{figure}[!ht]
\hspace{2.5cm}
\includegraphics[width=0.5\textwidth]{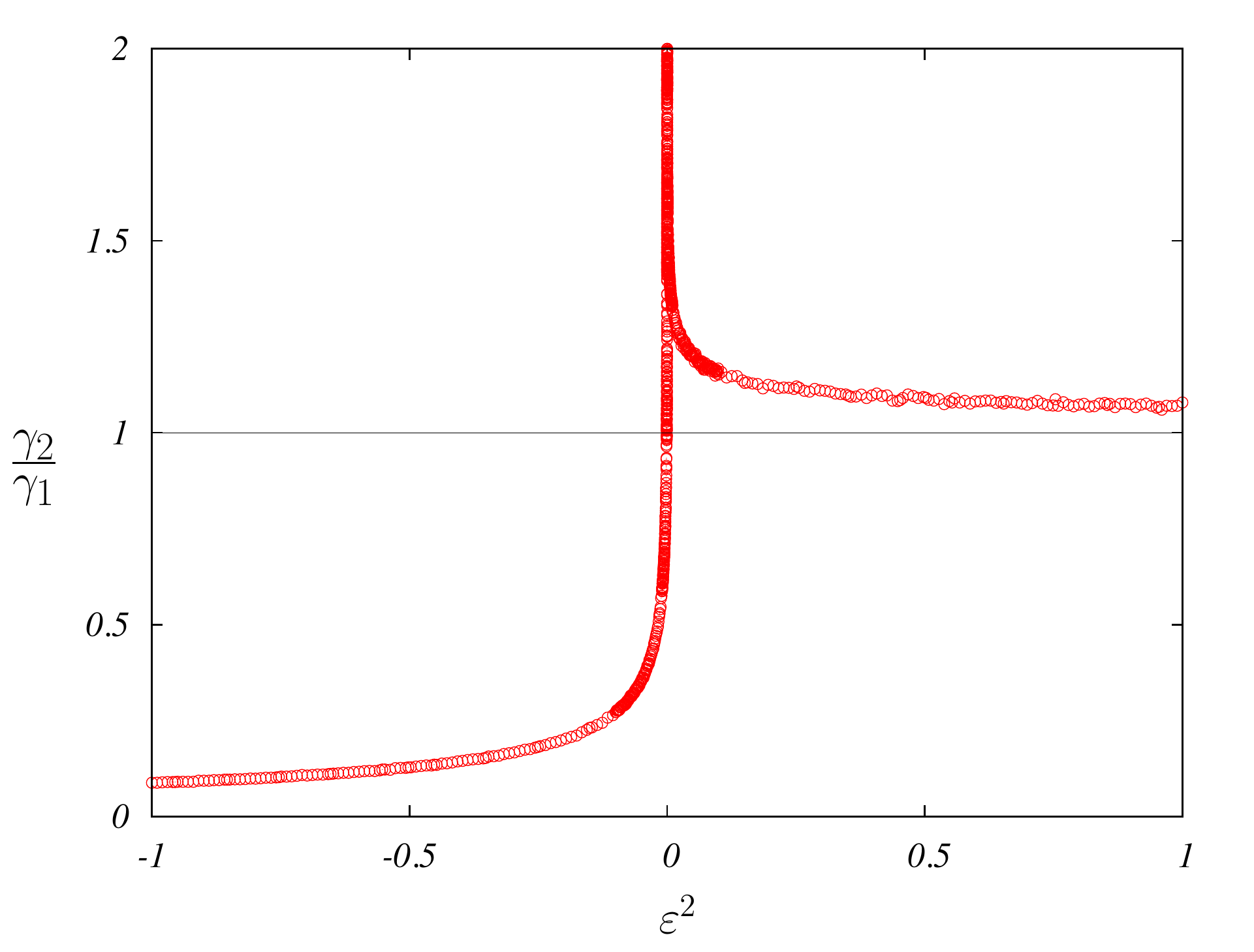}
\caption{\it (color online) The ratio of the two first cumulants for the random mass Dirac model presents a logarithmic divergence for $|\varepsilon|\to0$.}
\label{fig:anomaly}
\end{figure}

The standard weak disorder expansion for the tight-binding (Anderson) model
\begin{equation}
\label{eq:AndersonModel}
-t\,\varphi_{n+1}+V_n\,\varphi_n-t\,\varphi_{n-1}=\varepsilon\,\varphi_n
\end{equation}
is known to break down at the band center ($\varepsilon=0$) \cite{CzyKraMac81,KapWeg81}.
Whereas the standard expansion gives \cite{DerGar84,Luc92}
$\gamma_1\simeq a\,\mean{V_n^2}/(8t^2\sin^2\kappa)$ 
at $\varepsilon=-2t\cos\kappa$ for uncorrelated potentials $V_n$ (a is the lattice spacing),
the correct behaviour in the band center is 
$\gamma_1=a\,[\Gamma(3/4)/\Gamma(1/4)]^2\mean{V_n^2}/t^2+\mathcal{O}(\varepsilon)$ 
\cite{DerGar84}.
This small difference, $0.125$ vs $0.114..$, and those of other physical quantities, have been referred to as band center ``anomalies''~\footnote{
As shown in Ref.~\cite{DerGar84}, the occurence of anomalies is not specific to the band center but is an effect of commensurability.
}.
This phenomenon may be easily analysed within our continuum description~:
the continuum limit of the Anderson model \eqref{eq:AndersonModel} near the band center ($\varepsilon\to0$) is the random Dirac equation~\cite{Gog82}
\begin{equation}
[-\I\sigma_3\,\partial_x+V_0(x)+\sigma_1\,V_\pi(x)]\,\widetilde{\Psi}(x)
=\varepsilon\,\widetilde{\Psi}(x)
\end{equation}
(for $2at=1$), where $V_0(x)$ and $V_\pi(x)$ describe forward and backward (umklapp) scattering, respectively.
This makes it clear that the disorder cannot be treated perturbatively for $\varepsilon=0$.
After a rotation 
\begin{equation}
\Psi=\frac{1}{\sqrt2}(1-\I\sigma_1)\,\widetilde{\Psi}
\end{equation}
and choosing $V_0(x)=\sum_nv_n\,\delta(x-x_n)$ and $V_\pi(x)=\sum_n\eta_n\,\delta(x-x_n)$, this disordered model can be described by transfer matrices \eqref{eq:MSusy1}, by setting the angles $\theta_n=\varepsilon\,(x_{n+1}-x_n)-v_n$.
For $\varepsilon=0$, the Lyapunov exponent in the continuum limit is expressed in terms of elliptic integrals (this case was considered in section 6 of Ref.~\cite{ComLucTexTou13})~:
\begin{equation}
  \label{eq:LyapunovRandomAnglesEtas}
  \clp_1 = g\,
  \left[
    \frac{1}{k^2}\left(\frac{\mathbf{E}(k)}{\mathbf{K}(k)}-1\right)+1
  \right]
  \hspace{0.25cm}\mbox{ with }
  k=\frac{1}{\sqrt{1+g_0/g}}
\end{equation}
where $g=\rho\mean{\eta_n^2}$ and $g_0=\rho\mean{v_n^2}$.
For uncorrelated site potentials, $\mean{V_nV_m}\propto\delta_{n,m}$, we have $g_0=g$~; Eq.~\eqref{eq:LyapunovRandomAnglesEtas} with $k=1/\sqrt{2}$ leads to $\clp_1 = g\,[2\Gamma(3/4)/\Gamma(1/4)]^2$, in perfect correspondence with the result of Ref.~\cite{DerGar84}. 
$\clp_2$ at the band center was found in Ref.~\cite{SchTit03} where it was shown that the anomaly is \textit{small}, $\clp_2/\clp_1\simeq1.047$.
On the other hand, the suppression of forward scattering~\footnote{
  In the Anderson model, forward and backward scattering may be adjusted as follows~:
  one considers random potentials $V_n=V_0(na)+(-1)^n\,V_\pi(na)$, where $V_0(x)$ and $V_\pi(x)$ are two independent random functions varying smoothly at the scale of the lattice spacing $a$.
  Forward scattering is controlled by the strength $g_0$ of $V_0(x)$ whereas backward scattering is due to anti-correlation of nearest neighbour potentials, described by $V_\pi(x)$ with strength $g$.
} 
($g_0\ll g$) leads to the model studied in the present paper with a \textit{strong} anomaly $\clp_2/\clp_1\sim\ln(g/g_0)$ at $\varepsilon=0$ and  $\clp_2/\clp_1\sim\ln(g/|\varepsilon|)$ for $g_0\ll|\varepsilon|\ll g$, Fig.~\ref{fig:anomaly} (the value of $\clp_2$ for finite $g_0\ll g$ is deduced from a continuity assumption). 
Our continuous description thus makes clear how one can tune the band center anomaly by adjusting the relative magnitude of forward and backward scattering.

\section{Conclusion}

In this paper we have characterized the statistical properties of random matrix products 
$\Pi_N=M_N\cdots M_2M_1$ for two subgroups of $\mathrm{SL}(2,\mathbb{R})$, by making use of the fact that, for a certain choice of the distribution of the angles in \eqref{eq:MSusy1} and \eqref{eq:MSusy2}, $\ln||\Pi_N||$ can be simply expressed in terms of a Markov process~\cite{ComLucTexTou13}.
We have deduced the variance explicitly~; the integral representations Eqs.~(\ref{eq:TheResultSusy},\ref{eq:Gamma2Susy}) were demonstrated to be convenient for extracting limiting behaviours.
Following Ref.~\cite{PalVul87} and making use of \eqref{eq:UsefulRelation}, the generalised Lyapunov exponent \eqref{eq:DefGeneralisedLyapunov} may be obtained as the largest eigenvalue of the operator 
$\mathscr{G}^\dagger+q\,(z-\varepsilon^2/z)/2$.
The cumulants can be obtained by using the perturbative method used in Refs.~\cite{SchTit02,SchTit03}, however, apart from $\clp_1$, this leads to integral representations that seem less convenient to handle.
We have also obtained an integral representation similar to \eqref{eq:TheResultSusy} in the case of two other particular subgroups of $\mathrm{SL}(2,\mathbb{R})$ (see Appendix~\ref{appendix:Schrod}), corresponding to the model studied in Ref.~\cite{SchTit02}.
It remains a challenging issue to obtain a resolution of this problem, in the spirit of the general classification of Lyapunov exponents provided recently in~Ref.~\cite{ComLucTexTou13}.

\section*{Acknowledgements}

We acknowledge stimulating discussions with Alain Comtet, Bernard Derrida, Thierry Jolicoeur and Satya Majumdar, and a helpful suggestion of Anupam Kundu.

\appendix

\section{Details of the derivation of Eq.~\eqref{eq:Gamma2Susy}}
\label{appendix:DetailsOnCentralEq}

\subsection{$\varepsilon^2>0$}

For real energy, we see from the SDE \eqref{eq:SDEsusy} that the process $z(x)$ flows towards $\mathbb{R}$.
We have introduced the change of variable $\zeta=\mp\ln(\pm z/|\varepsilon|)/2$ for $z\in\mathbb{R}_\pm$, implying that the new process $\zeta(x)$ crosses $\mathbb{R}$ twice when $z(x)$ does once.
Hence the change of variable maps the SDE  \eqref{eq:SDEsusy} onto the couple of SDEs 
\begin{equation}
 \label{eq:LangevinZeta2}
  \deriv{}{x}\zeta 
  = |\varepsilon|\,\cosh2\zeta \pm \mass(x)
  = -\mathcal{U}_\pm'(\zeta) + \sqrt{g}\,\eta(x)
  \:.
\end{equation}
In the main text we used $\mean{\mass(x)}=0$ and the local nature of the mass correlation to disregard the sign.
Here we consider for the moment the general case $\mean{\mass(x)}=\mu\,g$ and introduce a couple of potentials $\mathcal{U}_\pm(\zeta)=-(|\varepsilon|/2)\sinh2\zeta\mp\mu\zeta$ related to the cases $z(x)>0$ and $z(x)<0$, respectively.
$\eta(x)$ is a normalised Gaussian white noise with zero mean.
The process is characterised by two stationary distributions $\mathcal{P}^\pm(\zeta)$, each normalized, related to $f(z)$ for $z\in\mathbb{R}_\pm$.
For example, the Lyapunov exponent is given by~\cite{ComTexTou10}
\begin{align}
  \clp_1 = \mean{z+\mass(x)}
  = \mu g+ \frac{|\varepsilon|}{2}
  \left[
    \int\D\zeta\,\mathcal{P}^+(\zeta)\,\EXP{-2\zeta}
    -\int\D\zeta\,\mathcal{P}^-(\zeta)\,\EXP{+2\zeta}
  \right]
  \:.
\end{align}
Now considering the case $\mu=0$ for which $\mathcal{P}^+=\mathcal{P}^-$, leads to 
$\clp_1 = -|\varepsilon|\,\mean{\sinh2\zeta}$, i.e.
\begin{equation}
  \label{eq:LyapZetaSM}
  \clp_1 = 2\mean{\mathcal{U}(\zeta)}
  \:.
\end{equation}
Fluctuations may be discussed in a similar way.
A crucial observation is that, in the original SDE \eqref{eq:SDEsusy}, the diffusion effectively vanishes at $z=0$, implying the absence of correlations between the process at coordinates $x$ and $x'$ associated to $z(x)>0$ and $z(x')<0$. 
It follows that the contributions of the fluctuations related to the two intervals $z\in\mathbb{R}_+$ and $z\in\mathbb{R}_-$ simply add.
The second term of \eqref{eq:TheResultSusy} takes the form
\begin{align}
\frac{1}{2}\bigg(
  &\int\D\zeta\D\zeta'\, |\varepsilon|\EXP{-2\zeta} \mathcal{G}^+(\zeta|\zeta')\,
  (-2|\varepsilon|)\sinh2\zeta'\, \mathcal{P}^+(\zeta')
  \nonumber\\\nonumber 
&+  \int\D\zeta\D\zeta'\, (-|\varepsilon|)\EXP{+2\zeta} \mathcal{G}^-(\zeta|\zeta')\,
  (-2|\varepsilon|)\sinh2\zeta'\, \mathcal{P}^-(\zeta') 
\bigg)
\:.
\end{align}
For $\mu=0$ we have $\mathcal{G}^+=\mathcal{G}^-$ leading to Eq.~\eqref{eq:Gamma2Susy}.

\subsection{$\varepsilon^2<0$}

For imaginary energy the analysis is slightly different~:
the process $z(x)$ is trapped on $\mathbb{R}_+$ and $\zeta(x)$ does not flow across $\mathbb{R}$.
The change of variable is simply $\zeta=-(1/2)\ln(z/|\varepsilon|)$. The new process is trapped by the potential well $\mathcal{U}(\zeta)=(|\varepsilon|/2)\cosh2\zeta-\mu\zeta$.
The equilibrium distribution is $\mathcal{P}(\zeta)\propto\exp\big[-(2/g)\mathcal{U}(\zeta)\big]$.
When $\mu=0$ the potential is symmetric. We can symmetrize the expression $\clp_1=|\varepsilon|\mean{\EXP{-2\zeta}}$, leading to $\clp_1=|\varepsilon|\mean{\cosh2\zeta}$, i.e. again to \eqref{eq:LyapZetaSM}.
Eq.~\eqref{eq:TheResultSusy} leads to
\begin{align}
  \label{eq:IntermediateGamma2}
  \clp_2 = g &+ 2|\varepsilon|\mean{\zeta\,\EXP{-2\zeta}}
  + 2|\varepsilon|^2
  \int\D\zeta\D\zeta'\,
     \EXP{-2\zeta}\,  \mathcal{G}(\zeta|\zeta')\, \cosh2\zeta' \,\mathcal{P}(\zeta')
     \:.
\end{align}
The second term can be obviously symmetrized, which gives the second term of \eqref{eq:Gamma2Susy}.
Symmetrization of the third integral term works as follows~:
the propagator may be decomposed over the left/right eigenvectors of the forward generator $\mathscr{G}^\dagger$ as 
\begin{equation}
  \mathcal{G}(\zeta|\zeta') = \sum_{n>0} 
  \frac{\Phi_n^\mathrm{R}(\zeta)\Phi_n^\mathrm{L}(\zeta')}{\mathscr{E}_n}
\end{equation}
where 
$\mathscr{G}^\dagger \Phi_n^\mathrm{R}(\zeta) = -\mathscr{E}_n\Phi_n^\mathrm{R}(\zeta)$
and 
$\mathscr{G} \Phi_n^\mathrm{L}(\zeta) = -\mathscr{E}_n\Phi_n^\mathrm{L}(\zeta)$.
Because the potential $\mathcal{U}(\zeta)$ is symmetric, the eigenvectors have a symmetry property 
$\Phi_n^\mathrm{L/R}(-\zeta)=(-1)^n\Phi_n^\mathrm{L/R}(\zeta)$.
Integration over $\zeta'$ in \eqref{eq:IntermediateGamma2} selects only the contributions of even eigenvectors which allows one to symmetrize the integrand with respect to $\zeta\to-\zeta$, leading to Eq.~\eqref{eq:Gamma2Susy}.

It is remarkable that despite the dynamics of the process $\zeta(x)$ being quite different for real and imaginary $\varepsilon$, we have found a unique representation for both $\clp_1$, Eq.~\eqref{eq:EpressionLyapunov}, and $\clp_2$, Eq.~\eqref{eq:Gamma2Susy}, expressed in terms of the potential $\mathcal{U}(\zeta)$.

\section{Direct calculation of $\clp_2$ for $\varepsilon=0$}
\label{appendix:FlucZero}

The study of the case $\varepsilon=0$ shows some subtlety related to the choice of the norm of the matrix.
In the usual case, the statistical properties of the RMP are independent of the precise definition of the norm~\cite{BouLac85,CriPalVul93}.
Bougerol and other authors propose
\begin{equation}
  \label{eq:Bougerol}
  ||M|| = \mathrm{Sup}\{ |Mx| \ ;\ x\in\mathbb{R}^2\ ;\ |x|=1\}
\end{equation}
where $|x|$ is the norm on the vector space.

In the numerical calculation, we have parametrized the spinor as $\Psi=\EXP{\xi}(\sin\Theta,-\cos\Theta)$, in the spirit of the phase formalism \cite{LifGrePas88}, and study the statistical properties of $\xi(x)=(1/2)\ln\big[\Psi(x)^\dagger\Psi(x)\big]$, usually setting $\Theta(0)=\Theta_0=0$.
Let us discuss the general case where $\Theta_0$ may differ from $0$.
Since $\Psi(x_{N+1}^-)=\Pi_N\Psi(x_1^-)$, the numerical procedure corresponds to considering the norm
\begin{equation}
 ||\Pi_N||_{\Psi_0} = |\Pi_N\Psi_0|
   \hspace{0.5cm}
  \mbox{ with }
  \Psi_0=
  \left(
  \begin{array}{c}
    \phantom{-}\sin\Theta_0 \\ -\cos\Theta_0
  \end{array}    
  \right)
  \:,
\end{equation}
i.e. $\xi(x_{N+1}^-)=\ln||\Pi_N||_{\Psi_0}$.
We also introduce another possible definition of the norm 
\begin{equation}
  |||\Pi_N||| = \int_{|\Psi_0|=1}\D\Psi_0\, ||\Pi_N||_{\Psi_0}
  \:,
\end{equation}
closer to the spirit of \eqref{eq:Bougerol}.

For $\varepsilon=0$, the matrix product $\Pi_N$ can be studied rather directly~:
the angles vanish $\theta_n=0$ and the matrices $M_n$ commute. Hence we can write
\begin{equation}
  \Pi_N = 
  \left(
  \begin{array}{cc}
    \EXP{\Lambda} & 0 \\
    0 & \EXP{-\Lambda}
  \end{array}    
  \right)
  \hspace{0.5cm}
  \mbox{ with }
  \Lambda = \sum_{n=1}^N\eta_n
  \:.
\end{equation}
The distribution of the random variable $\Lambda$ is given by the central limit theorem~:  $\mean{\Lambda}=\rho x\mean{\eta_n}=0$ and 
$\mathrm{Var}(\Lambda)=\rho x\mean{\eta_n^2}=gx$
(we consider that $x$ is fixed and $N$ fluctuates with $\mean{N}=\rho x$).
We have 
\begin{equation}
  \label{eq:Norm1ForZeroEnergy}  
  \ln||\Pi_N||_{\Psi_0} = 
  \frac12 \ln\left[\cosh2\Lambda-\cos2\Theta_0\,\sinh2\Lambda\right]
  \:.
\end{equation}

We examine first the particular case $\Theta_0=0$, leading to $\ln||\Pi_N||_{\Psi_0} = -\Lambda$.
We immediatly deduce that 
$\mean{\ln||\Pi_N||_{\Psi_0}}=0$
and 
$\mathrm{Var}(\ln||\Pi_N||_{\Psi_0})=gx$, which would lead to $\clp_1=0$ and, incorrectly, to $\clp_2=g$.
The choice $\Theta_0=\pi/2$ leads to a similar conclusion.
This reflects the statistical properties of the two particular zero energy solutions 
\begin{equation}
  \left(\begin{array}{c} 1 \\ 0 \end{array}\right)\,\EXP{\int^x\D x'\,\mass(x')}
  \hspace{0.25cm}\mbox{ and }\hspace{0.25cm}
  \left(\begin{array}{c} 0 \\ 1 \end{array}\right)\,\EXP{-\int^x\D x'\,\mass(x')}
\end{equation}
selected by the choices $\Theta_0=\pi/2$  and $\Theta_0=0$, respectively.

We now consider the case of an arbitrary initial vector, with $\Theta_0\notin\{0,\,\pi/2\}$.
In the $N\to\infty$ limit, the large $\Lambda$ behaviour of the norm is selected~: 
$
\ln||\Pi_N||_{\Psi_0} \simeq 
|\Lambda| +\heaviside(\Lambda)\,\ln|\sin\Theta_0|
+\heaviside(-\Lambda)\,\ln|\cos\Theta_0|
$.
Some algebra gives, for $gx\gg1$,
\begin{equation}
  \label{eq:MeanLogPi}
  \mean{ \ln||\Pi_N||_{\Psi_0} } 
  \simeq \sqrt{\frac{2gx}{\pi}} + \frac{1}{2}\ln\left|\frac12\sin2\Theta_0\right|
\end{equation}
and 
\begin{equation}
  \label{eq:VarLogPi}
  \mathrm{Var}(\ln||\Pi_N||_{\Psi_0})
  \simeq gx\left(1-\frac{2}{\pi}\right) + \frac14\ln^2|\tan\Theta_0|  
  \:.
\end{equation}
Note that the average value is reminiscent of the average of the logarithm of the transmission probability \cite{SteCheFabGog99} (this calculation was first performed in Ref.~\cite{ComMonYor98} in another context).
Interestingly, the behaviours (\ref{eq:MeanLogPi},\ref{eq:VarLogPi}) were shown to persist in a quasi-one-dimensional situation with an \textit{odd} number of channels (see the review~\cite{BroMudFur01} and references therein).
We obtain 
\begin{align}
  \label{eq:DirectGamma1}
  \clp_1 &= 0 \\
  \label{eq:DirectGamma2}
  \clp_2 &= g\left(1-\frac{2}{\pi}\right)= g\times0.363380...
\end{align}

We can easily repeat this calculation with the second norm. Averaging of \eqref{eq:Norm1ForZeroEnergy} over the angle $\Theta_0$ gives 
\begin{equation}
  \label{eq:Norm2ForZeroEnergy}  
  |||\Pi_N||| = 
  \frac{2\EXP{|\Lambda|}}{\pi}\,
  \mathbf{E}\left(\sqrt{1-\EXP{-4|\Lambda|}}\right)
  \:,
\end{equation}
where $\mathbf{E}(k)$ is the elliptic integral~\cite{gragra}.
We deduce the asymptotic behaviours 
$\ln|||\Pi_N|||\simeq(3/4)\Lambda^2$ for $|\Lambda|\ll1$ and 
$\ln|||\Pi_N|||\simeq|\Lambda|-\ln(\pi/2)$ for $|\Lambda|\gg1$, leading again to  (\ref{eq:DirectGamma1},\ref{eq:DirectGamma2}).

In conclusion~:
for $\varepsilon\neq0$, the calculation of the cumulants $\clp_n$ is insensitive to the precise definition of the norm, i.e. to the precise choice of the initial spinor. 
In the Monte Carlo simulation, we have chosen $\Theta_0=0$ in order to set a Dirichlet boundary condition for the first component of the spinor. 
On the other hand, setting $\varepsilon=0$, the behaviour of $\clp_2$ as a function of $\Theta_0$ presents two discontinuities precisely at $0$  and $\pi/2$. We understand these singular values as resulting from a lack of ergodicity in the matrix space when considering the Abelian subgroup describing the case $\varepsilon=0$. Hence, the value $g$ found for $\Theta_0=0$ or $\pi/2$ should not be taken as the correct result.


\section{Two other subgroups of random matrices of $\mathrm{SL}(2,\mathbb{R})$}
\label{appendix:Schrod}

It is well-known that the random Kronig-Penney model
$\big[-\partial_x^2 + \sum_n v_n \,\delta(x-x_n) \big]\psi(x) = E\,\psi(x)$ for energy $E=k^2$
is controlled by transfer matrices of the form 
\begin{equation}
  \label{eq:MSchrod1}
  M_n = 
  \left(
  \begin{array}{cc}
    \cos\theta_n & -\sin\theta_n \\
    \sin\theta_n & \phantom{-}\cos\theta_n 
  \end{array}    
  \right)
  \left(
  \begin{array}{cc}
    1 & u_n \\
    0 & 1
  \end{array}    
  \right)
\end{equation}
where $\theta_n=k\,(x_{n+1}-x_{n})>0$ and $u_n=v_n/k$.
The Schr\"odinger equation with negative energy $E=-k^2$ involves matrices of the form~\cite{ComTexTou10}
\begin{equation}
  \label{eq:MSchrod2}
  M_n = 
  \left(
  \begin{array}{cc}
    \cosh\theta_n & \sinh\theta_n \\
    \sinh\theta_n & \cosh\theta_n 
  \end{array}    
  \right)
  \left(
  \begin{array}{cc}
    1 & u_n \\
    0 & 1
  \end{array}    
  \right)
\end{equation}
with the same definitions for $\theta_n$ and $u_n$.

The study of the continuum limit, $\ell_n\to0$ and $v_n\to0$ with $\mean{v_n}=0$ and $\sigma=\mean{v_n^2}/\mean{\ell_n}$ fixed can be done along the same lines as in the paper.
In this more simple case, the Riccati variable $z(x)=\psi'(x)/\psi(x)$ obeys the SDE 
$z'(x)=-E-z(x)^2+V(x)$. In the continuum limit $V(x)$ is a Gaussian white noise of variance $\sigma$ and the process is characterised by the (backward) generator 
$\mathscr{G} = ({\sigma}/{2})\partial_z^2-(E+z^2)\partial_z$. 
We arrive at
\begin{equation}
  \label{eq:VarianceSchrodinger}
  \clp_2 = 2 \,\int\D z\D z'\, z\, G(z|z') \, z'\, f(z')
\end{equation}
where 
\begin{equation}
  f(z) = \frac{2N}{\sigma}
  f_0(z)
  \int_{-\infty}^z\frac{\D t}{f_0(t)}
  \hspace{0.25cm}\mbox{with }
  f_0(z)=\EXP{-\frac{2}{\sigma}\mathcal{U}(z)}
\end{equation}
is the stationary distribution, involving the potential $\mathcal{U}(z) = Ez+(1/3)z^3$ and the integrated density of states $N(E)$, given in Ref.~\cite{Hal65} for instance (also recalled in Ref.~\cite{GraTexTou14}).
The equation
\begin{equation}
  \label{eq:EqForG}
  \mathscr{G}^\dagger  G(z|z') = f(z) - \delta(z-z') 
\end{equation}
for the propagator can be solved~:
\begin{equation}
  G(z|z')= \frac{1}{N(E)}
  \left\{
    f(z) \left[c(z')+\int_{-\infty}^z\D t\,f(t)\right]
    - f_0(z)  \int_{-\infty}^z\D t\,\frac{f(t)^2}{f_0(t)}
    + \frac{f_0(z_>)f(z_<)}{f_0(z')} 
  \right\}
\end{equation}
where 
\begin{equation}
  c(z') +\frac12 = 
  \frac{\sigma}{2N(E)}
  \left[\int_{-\infty}^{+\infty}\D z'' \, f(z'')^2 f(-z'')
  -f(-z')\,f(z')
  \right]
  - \int_{-\infty}^{z'} \D z''\, f(z'')
  \:.
\end{equation}

We can analyse the limiting behaviours of the variance \eqref{eq:VarianceSchrodinger}.
In the high energy regime, $k=\sqrt{E}\gg\sigma^{1/3}$ we obtain the expansions
\begin{equation}
  f(z) = \frac{k/\pi}{z^2+k^2} + \frac{\sigma k}{\pi}\frac{z}{(z^2+k^2)^3}  + \mathcal{O}(\sigma^2)
\end{equation}
(recall that $N(E)={k}/{\pi}+\mathcal{O}(\sigma^2)$) and
\begin{align}
  \label{eq:ExpansionG}
  G(z|z')
  =&\left[
  \frac{1}{z^2+k^2} 
  +\sigma \frac{z}{(z^2+k^2)^3}
  \right]\Omega(z,z')
  \\\nonumber
  &+\frac{3\sigma}{16\pi k^3}
  \left( \frac{1}{z^2+k^2}-\frac{4k^4}{(z^2+k^2)^3}\right)
  +
  \frac{\heaviside(z-z')}{z'^2+k^2}\frac{f_0(z)}{f_0(z')}  
  +\mathcal{O}(\sigma^2)
\end{align}
where 
\begin{equation}
  \Omega(z,z') =
  \frac12\sign(z'-z)+\frac{1}{\pi}
  \big[\arctan(z/k)-\arctan(z'/k)\big]
\:.
\end{equation}
When introducing these expressions in \eqref{eq:VarianceSchrodinger}, the term $\mathcal{O}(\sigma^0)$ seems at first sight logarithmically divergent but is eliminated by symmetry (i.e. integrals must be understood as principal parts).
We get
\begin{equation}
  \label{eq:Kappa2SchrodEnergyPos}
  \clp_2 = \frac{k\sigma}{\pi} \int \D z\,
  \frac{z^2}{[\mathcal{U}'(z')]^3}  
  +\mathcal{O}(\sigma^2)
  =
  \frac{\sigma}{8E}+\mathcal{O}(\sigma^2)
\end{equation}
i.e. we have recovered the asymptotic relation $\clp_2\simeq\clp_1$ for $E\to\infty$ (SPS).

\begin{figure}[!ht]
\includegraphics[width=0.5\textwidth]{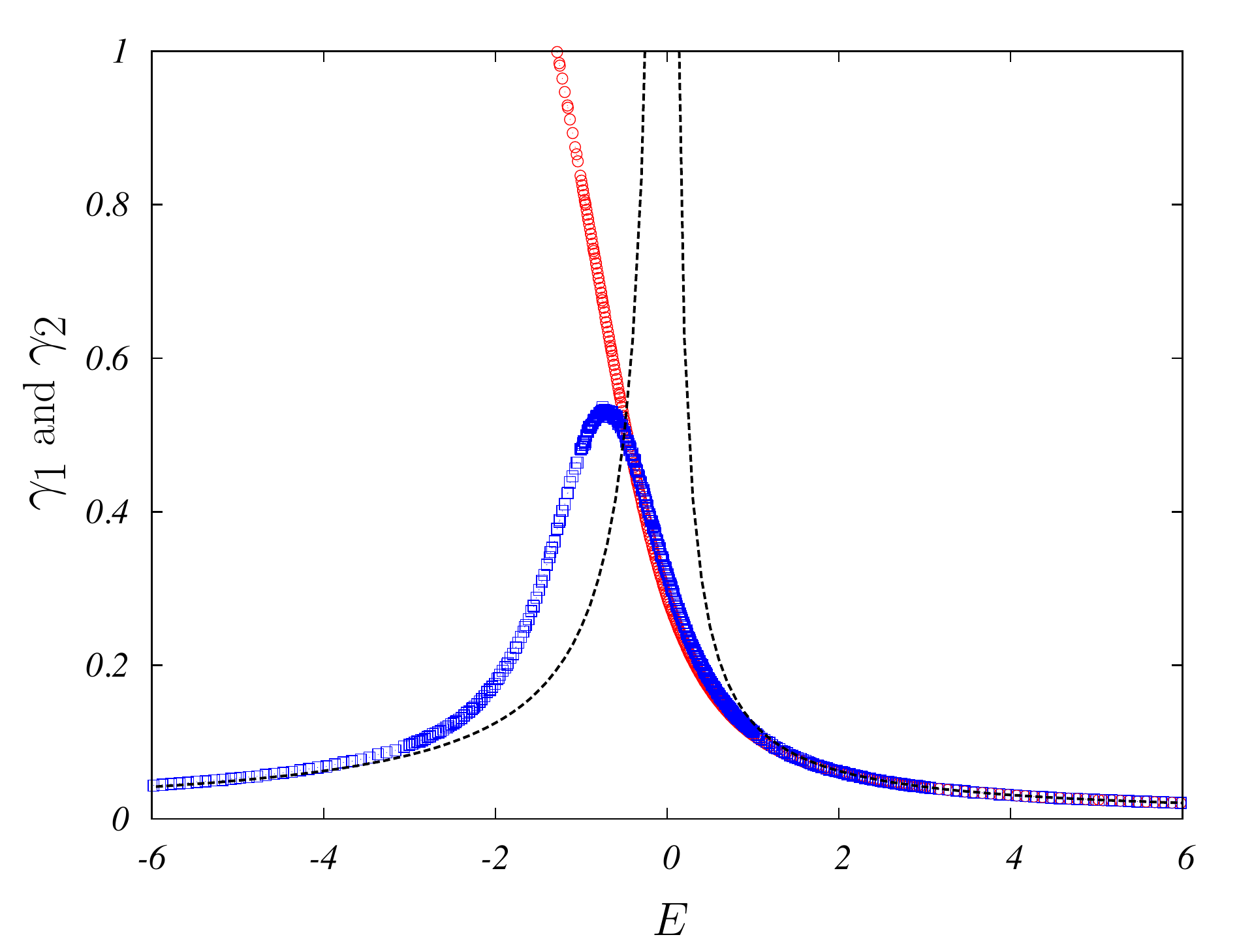}
\hfill
\includegraphics[width=0.5\textwidth]{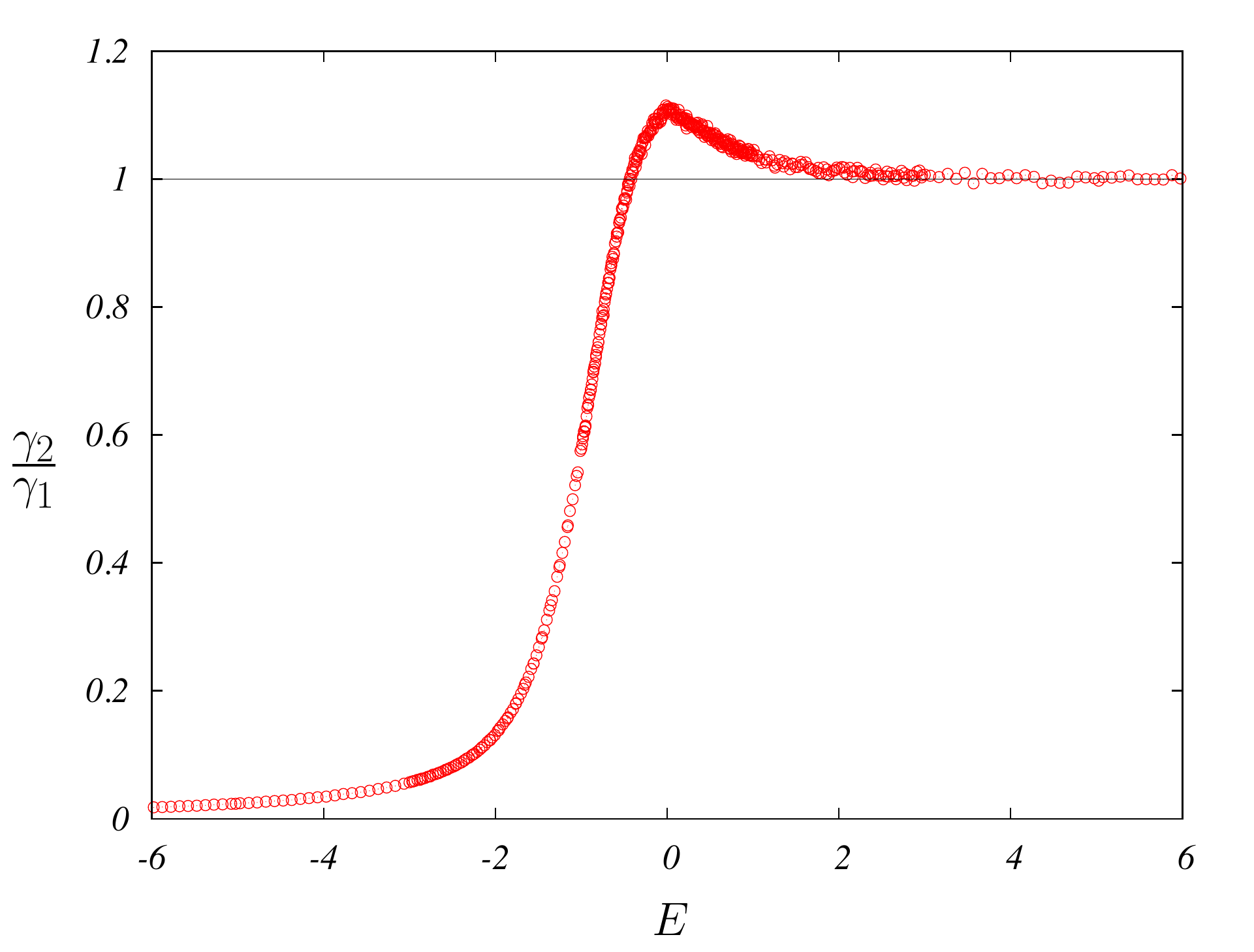}
\caption{\it (color online) 
Left~: Plot of the Lyapunov exponent (red circles) and the variance (blue squares) for $\sigma=1$ obtained by Monte Carlo simulations. Comparison with limiting behaviours \eqref{eq:Kappa2SchrodEnergyPos} and \eqref{eq:Kappa2SchrodEnergyNeg} (dashed black lines).
Right~: SPS, $\clp_2/\clp_1\simeq1$,  holds for $E\gg\sigma^{2/3}$.}
\label{fig:gamma2Schrod}
\end{figure}

For $E=0$, the fluctuations are finite $\clp_2=\tilde{c}\,\sigma^{1/3}$ where $\tilde{c}$ is a dimensionless constant of order unity (calculated explicitly in Ref.~\cite{SchTit02}).
$\clp_2$ is maximum for a negative value of the energy, however the numerics shows that the ratio $\clp_2/\clp_1$ reaches its maximum at $E=0$ (Fig.~\ref{fig:gamma2Schrod}).

The limit $k=\sqrt{-E}\gg\sigma^{1/3}$ is more easy to handle.
In this case the potential $\mathcal{U}(z)$ develops a deep well at $z=k$, where the process is most of the ``time'' trapped. This dominates the fluctuations, which are those of the Ornstein-Uhlenbeck process,
\begin{equation}
  \label{eq:Kappa2SchrodEnergyNeg}
  \clp_2   
  \underset{E\to-\infty}{\simeq}
 \frac{\sigma}{4(-E)}
  \:.
\end{equation}
The fluctuations thus decay \textit{faster} as energy decreases than in the Dirac case studied in the paper, since the relation between the two models involves the mapping $E\leftrightarrow\varepsilon^2$.
Recalling that $\clp_1\simeq\sqrt{-E}$ in this case shows that $\clp_2\ll\clp_1$ (no SPS).

Monte Carlo simulations are in perfect agreement with these behaviours (see Fig.~\ref{fig:gamma2Schrod}).

The problem considered in this appendix was studied earlier in Refs.~\cite{SchTit02,ZilPik03} in another context and with a different method~:
the generalised Lyapunov exponent \eqref{eq:DefGeneralisedLyapunov} is obtained as the largest eigenvalue of the operator $\mathscr{G}^\dagger +qz$ \cite{PalVul87}.
The perturbative treatment \cite{SchTit02} gives an integral representation 
\begin{equation}
  \label{eq:SchomerusTitov2002}
  \clp_2 = 2 \int\D z \, (z-\clp_1)\, \varphi_1(z)  
\end{equation}
where 
\begin{align}
  \varphi_1(z) = N\left(\frac{2}{\sigma}\right)^2f_0(z)
  \int_{-\infty}^z\frac{\D z'}{f_0(z')} 
  \int_{-\infty}^{z'}\D z''\,(\clp_1-z'')\,f_0(z'')
  \int_{-\infty}^{z''}\frac{\D z'''}{f_0(z''')} 
  \:.
\end{align}
Although it is not straightforward to prove the equivalence between \eqref{eq:VarianceSchrodinger} and \eqref{eq:SchomerusTitov2002}, they seem to give similar results (see Fig.~1 of Ref.~\cite{SchTit02}).


\addcontentsline{toc}{section}{\protect\bibname}
\begin{thebibliography}{10}

\bibitem{Bax82}
R.~J. Baxter,
  {\it Exactly Solved Models in Statistical Mechanics},
  Academic Press, London, 1982.

\bibitem{BouLac85}
P.~Bougerol and J.~Lacroix,
  {\it Products of Random Matrices with Applications to Schr\"{o}dinger
  Operators},
  Birkha\"{u}ser, Basel, 1985.

\bibitem{Luc92}
J.-M. Luck,
  {\it Syst\`emes d\'esordonn\'es unidimensionnels},
  CEA, collection Al\'ea Saclay, Saclay, 1992.

\bibitem{CriPalVul93}
A.~Crisanti, G.~Paladin, and A.~Vulpiani,
  {\it Products of random matrices in statistical physics},
  Springer-Verlag, 1993,
  Springer Series in Solid-State Sciences vol.~{\bf104}.

\bibitem{FigMosKno98}
M.~T. Figge, M.~V. Mostovoy, and J.~Knoester,
  Critical temperature and density of spin flips in the anisotropic
  random-field Ising model,
  Phys. Rev. B {\bf 58}, 2626--2634 (1998).

\bibitem{PalVul87b}
G.~Paladin and A.~Vulpiani,
  Anomalous scaling and generalized Lyapunov exponents of the
  one-dimensional Anderson model,
  Phys. Rev. B {\bf 35}, 2015--2020 (1987).

\bibitem{PalVul87}
G.~Paladin and A.~Vulpiani,
  Anomalous scaling in multifractal objects,
  Phys. Rep. {\bf 156}(4), 147--225 (1987).

\bibitem{AltPri89}
B.~L. Altshuler and V.~N. Prigodin,
  Distribution of local density of states and NMR line shape in a
  one-dimensional disordered conductor,
  Sov. Phys. JETP {\bf 68}(1), 198--209 (1989).

\bibitem{TexCom99}
C.~Texier and A.~Comtet,
  Universality of the Wigner time delay distribution for
  one-dimensional random potentials,
  Phys. Rev. Lett. {\bf 82}(21), 4220--4223 (1999).

\bibitem{AndThoAbrFis80}
P.~Anderson, D.~J. Thouless, E.~Abrahams, and D.~S. Fisher,
  New method for a scaling theory of localization,
  Phys. Rev. B {\bf 22}(8), 3519--3526 (1980).

\bibitem{CohRotSha88}
A.~Cohen, Y.~Roth, and B.~Shapiro,
  Universal distributions and scaling in disordered systems,
  Phys. Rev. B {\bf 38}(17), 12125--12132 (1988).

\bibitem{DeyLisAlt00}
L.~I. Deych, A.~A. Lisyansky, and B.~L. Altshuler,
  Single Parameter Scaling in One-Dimensional Localization Revisited,
  Phys. Rev. Lett. {\bf 84}(12), 2678 (2000).

\bibitem{SchTit03}
H.~Schomerus and M.~Titov,
  Band-center anomaly of the conductance distribution in
  one-dimensional Anderson localization,
  Phys. Rev.~B {\bf 67}, 100201 (2003).

\bibitem{TitSch05}
M.~Titov and H.~Schomerus,
  Nonuniversality of Anderson Localization in Short-Range Correlated
  Disorder,
  Phys. Rev. Lett. {\bf 95}, 126602 (2005).

\bibitem{ComLucTexTou13}
A.~Comtet, J.-M. Luck, C.~Texier, and Y.~Tourigny,
  The Lyapunov exponent of products of random $2\times2$ matrices close
  to the identity,
  J. Stat. Phys. {\bf 150}, 13--65 (2013).

\bibitem{ComTexTou10}
A.~Comtet, C.~Texier, and Y.~Tourigny,
  Products of random matrices and generalised quantum point scatterers,
  J. Stat. Phys. {\bf 140}(3), 427--466 (2010).

\bibitem{LeDMonFis99}
P.~{Le~Doussal}, C.~Monthus, and D.~S. Fisher,
  Random walkers in one-dimensional random environments: Exact
  renormalization group analysis,
  Phys. Rev.~E {\bf 59}(5), 4795 (1999).

\bibitem{TexHag10}
C.~Texier and C.~Hagendorf,
  Effect of boundaries on the spectrum of a one-dimensional random mass
  Dirac Hamiltonian,
  J.~Phys.~A: Math. Theor. {\bf 43}, 025002 (2010).

\bibitem{ComTex98}
A.~Comtet and C.~Texier,
  One-dimensional disordered supersymmetric quantum mechanics: a brief
  survey,
  in {\it Supersymmetry and Integrable Models}, edited by H.~Aratyn,
  T.~D. Imbo, W.-Y. Keung, and U.~Sukhatme, Lecture Notes in Physics, Vol. 502,
  pages 313--328, Springer, 1998, 
  (also available as cond-mat/97\,07\,313).

\bibitem{BouComGeoLeD90}
J.-P. Bouchaud, A.~Comtet, A.~Georges, and P.~{Le~Doussal},
  Classical diffusion of a particle in a one-dimensional random force
  field,
  Ann. Phys. (N.Y.) {\bf 201}, 285--341 (1990).

\bibitem{ComTexTou13}
A.~Comtet, C.~Texier, and Y.~Tourigny,
  Lyapunov exponents, one-dimensional Anderson localisation and
  products of random matrices,
  J.~Phys.~A: Math. Theor. {\bf 46}, 254003 (2013).

\bibitem{Gar89}
C.~W. Gardiner,
  {\it Handbook of stochastic methods for physics, chemistry and the
  natural sciences},
  Springer, 1989.

\bibitem{Tex99}
C.~Texier,
  {\it Quelques aspects du transport quantique dans les syst\`emes
  d\'esordonn\'es de basse dimension},
  PhD thesis, Universit\'e Paris 6, 1999,
  \url{http://lptms.u-psud.fr/christophe_texier/}.

\bibitem{BieTex08}
T.~Bienaim\'e and C.~Texier,
  Localization for one-dimensional random potentials with large
  fluctuations,
  J.~Phys.~A: Math. Theor. {\bf 41}, 475001 (2008).

\bibitem{GraTexTou14}
A.~Grabsch, C.~Texier, and Y.~Tourigny,
  One-dimensional disordered quantum mechanics and Sinai diffusion with
  random absorbers,
  J. Stat. Phys. {\bf 155}, 237--276 (2014).

\bibitem{LifGrePas88}
I.~M. Lifshits, S.~A. Gredeskul, and L.~A. Pastur,
  {\it Introduction to the theory of disordered systems},
  John Wiley \& Sons, 1988.

\bibitem{SteCheFabGog99}
M.~Steiner, Y.~Chen, M.~Fabrizio, and A.~O. Gogolin,
  Statistical properties of localization-delocalization transition in
  one dimension,
  Phys. Rev.~B {\bf 59}(23), 14848--14851 (1999).

\bibitem{SteFabGog98}
M.~Steiner, M.~Fabrizio, and A.~O. Gogolin,
  Random mass Dirac fermions in doped spin-Peierls and spin-ladder
  systems: one-particle properties and boundary effects,
  Phys. Rev.~B {\bf 57}(14), 8290--8306 (1998).

\bibitem{CzyKraMac81}
G.~Czycholl, B.~Kramer, and A.~MacKinnon,
  Conductivity and Localization of Electron States in One Dimensional
  Disordered Systems: Further Numerical Results,
  Z. Phys. B {\bf 43}, 5--11 (1981).

\bibitem{KapWeg81}
M.~Kappus and F.~Wegner,
  Anomaly in the band centre of the one-dimensional Anderson model,
  Z. Phys. B {\bf 45}(1), 15--21 (1981).

\bibitem{DerGar84}
B.~Derrida and E.~J. Gardner,
  Lyapounov exponent of the one dimensional Anderson model: weak
  disorder expansions,
  J. Physique {\bf 45}, 1283--1295 (1984).

\bibitem{Gog82}
A.~A. Gogolin,
  Electron localization and hopping conductivity in one-dimensional
  disordered systems,
  Phys. Rep. {\bf 86}(1), 1--53 (1982).

\bibitem{SchTit02}
H.~Schomerus and M.~Titov,
  Statistics of finite-time Lyapunov exponents in a random
  time-dependent potential,
  Phys. Rev. E {\bf 66}, 066207 (2002).

\bibitem{ComMonYor98}
A.~Comtet, C.~Monthus, and M.~Yor,
  Exponential functionals of Brownian motion and disordered systems,
  J. Appl. Probab. {\bf 35}, 255 (1998).

\bibitem{BroMudFur01}
P.~W. Brouwer, C.~Mudry, and A.~Furusaki,
  Transport properties and density of States of quantum wires with
  Off-Diagonal Disorder,
  Physica E {\bf 9}, 333--339 (2001).

\bibitem{gragra}
I.~S. Gradshteyn and I.~M. Ryzhik,
  {\it Table of integrals, series and products},
  Academic Press, fifth edition, 1994.

\bibitem{Hal65}
B.~I. Halperin,
  Green's Functions for a Particle in a One-Dimensional Random
  Potential,
  Phys. Rev. {\bf 139}(1A), A104--A117 (1965).

\bibitem{ZilPik03}
R.~Zillmer and A.~Pikovsky,
  Multiscaling of noise-induced parametric instability,
  Phys. Rev. E {\bf 67}, 061117 (2003).

\end{thebibliography}

\end{document}